\documentclass[11pt, a4paper]{article}
\usepackage{jheppub}

\newcommand {\cL}{{\cal L}}

\newcommand {\cN}{{\cal N}}
\newcommand {\cO}{{\cal O}}
\newcommand {\cP}{{\cal P}}

\def\G{\Gamma}

\def\q{\theta}

\def\z{\zeta}

\def\S{\Sigma}
\def\U{\Upsilon}

\def\ri{{\rm i}}

\newcommand{\ad}{{\dot{\alpha}}}

\newcommand{\dalpha}{{\dot{\alpha}}}

\newcommand{\1}{{\underline{1}}}
\newcommand{\2}{{\underline{2}}}

\newcommand{\pa}{\partial}

\newcommand{\be}{\begin{equation}}
\newcommand{\ee}{\end{equation}}
\newcommand{\bea}{\begin{eqnarray}}
\newcommand{\eea}{\end{eqnarray}}

\newcommand{\ba}{\begin{array}}
\newcommand{\ea}{\end{array}}

\def\double #1{#1{\hbox{\kern-2pt $#1$}}}

\newcommand{\bsubeq}{\begin{subequations}}
\newcommand{\esubeq}{\end{subequations}}

\newcommand{\rd}{\mathrm d}
\newcommand{\veps}{\varepsilon}

\newcommand{\abs}[1]{\left| #1 \right|}

\newcommand{\eol}{\notag \\}

\title{Relating harmonic and projective descriptions
of $\cN=2$ nonlinear sigma models}
\author{Daniel Butter}
\affiliation{School of Physics M013, The University of Western Australia \\
35 Stirling Highway, Crawley, W.A. 6009, Australia}
\emailAdd{daniel.butter@uwa.edu.au}

\abstract{Recent papers have established the relationship
between projective superspace and a complexified version of harmonic
superspace. We extend this construction to the case of general nonlinear
sigma models in both frameworks. Using an analogy with Hamiltonian
mechanics, we demonstrate how the Hamiltonian structure of the
harmonic action and the symplectic structure of the
projective action naturally arise from a single unifying action
on a complexified version of harmonic superspace. This
links the harmonic and projective descriptions of hyperk\"ahler
target spaces. For the two examples of Taub-NUT and Eguchi-Hanson,
we show how to derive the projective superspace solutions
from the harmonic superspace solutions.}

\begin{document}
\maketitle

\section{Introduction and review}

The relation between supersymmetry and complex geometry began with
Zumino's observation that $\cN=1$ supersymmetric nonlinear sigma models
must possess a K\"ahler geometry \cite{Zumino:1979et}. Soon after, it was discovered
that $\cN=2$ supersymmetry requires that the geometry be hyperk\"ahler \cite{A-GF}
(see also \cite{CF}).
However, while $\cN=1$ superspace naturally leads to
a real K\"ahler potential as the most general Lagrangian, the link between $\cN=2$ superspace
and hyperk\"ahler geometry is more complicated, due mainly to the complexity
of the superspaces that furnish off-shell representations of $\cN=2$ supersymmetry.

Harmonic superspace \cite{GIKOS, GIOS} extends the usual flat $\cN=2$ superspace by
an $S^2$ and deals with fields and operators which are globally defined
on this auxiliary manifold. Projective superspace \cite{KLR, LR:Projective, LR:SYM}
uses the same auxiliary manifold, though usually interpreted as
$\mathbb CP^1$, but requires fields and operators to be holomorphic
in the $\mathbb CP^1$ coordinate. The most general fields and actions
in projective superspace are necessarily singular somewhere on the auxiliary
manifold.

In 1998, Kuzenko showed that projective superspace could be understood
as a double-punctured harmonic superspace \cite{K:DP}. The main idea was
to embed projective multiplets into globally defined harmonic multiplets
that were holomorphic everywhere
except a small region around the poles of the $S^2$ (using the equation of
motion for the hypermultiplet and a gauge condition for the vector multiplet).
More recently, Jain and Siegel \cite{JS1} (see also \cite{JS2, JS3})
have constructed projective fields and actions from harmonic
ones using a different approach. Their idea is to complexify the $S^2$
and to interpret one of its dimensions as an additional coordinate which
can be integrated over once the equation of motion (for a hypermultiplet)
or a gauge condition (for a vector multiplet) is imposed.

Our goal in this paper is to take the same basic idea as \cite{JS1}
and apply it to the most general nonlinear sigma model
constructed from hypermultiplets in harmonic superspace. We will show that there exists a
natural identification with a model in projective superspace.
Moreover, once the nonlinear sigma model equations of motion are solved in
harmonic superspace, the corresponding projective superspace action
and solution can be derived. We will demonstrate the procedure for
three examples: the free hypermultiplet, Taub-NUT, and Eguchi-Hanson.

To keep this paper as self-contained as possible, we end this section with
a brief review of harmonic and projective superspaces and how both
describe nonlinear sigma models. These briefest of sketches are meant only to
remind the reader of the details and to clarify our conventions. We
also include a very brief review of an alternative action formulation
of Hamiltonian mechanics that will be necessary for our construction.

\subsection{Nonlinear sigma models in harmonic superspace}
Harmonic superspace \cite{GIKOS, GIOS} involves the extension of the global $\cN=2$ superspace
$\mathbb R^{4|8}$, parametrized by $z^M = (x^m, \q^\alpha_i, \bar\q_\dalpha^i)$,
with the two-sphere $S^2 \cong SU(2) / U(1)$, parametrized by the harmonics $u_i^\pm$.
The harmonics obey the relations
\begin{align}
\overline{u^{i +}} = u_i^-~, \qquad u^{i+} u_i^- = 1
\end{align}
and describe the space $SU(2)$.
Associated with the harmonics are three derivative operations
\begin{align}
D^{++} = u_i^+ \frac{\pa}{\pa u_i^-}~, \qquad
D^{--} = u^{i-} \frac{\pa}{\pa u^{i+}}~, \qquad
D^0 = u^{i+} \frac{\pa}{\pa u^{i+}} - u_i^- \frac{\pa}{\pa u_i^-}~,
\end{align}
corresponding to the three generators of $SU(2)$.
All superfields and operators in harmonic superspace are required to
possess a fixed $U(1)$ charge, that is, to be eigenstates of the
operator $D^0$. This allows the harmonics to be identified
modulo an overall phase,
\begin{align}\label{eq_HarmU1}
u_i^\pm \sim e^{\pm\ri \varphi} u_i^\pm~, \qquad \varphi \in \mathbb R~,
\end{align}
which ensures that only the coset space $SU(2) / U(1) \cong S^2$ is being
described.\footnote{To prove $SU(2) / U(1) \cong S^2$, introduce the coordinate
$X^I \in \mathbb R^3$ given by $X^I = u_j^- (\sigma^I)^j{}_k u^{k+}$
where $\sigma^I$ are the three Pauli matrices. One can check that
$\sum_I (X^I)^2 = 1$. Given any such $X^I$, one can reconstruct the
harmonics up to the equivalence relation \eqref{eq_HarmU1}.}

To construct an $\cN=2$ supersymmetric action, one needs
a Lagrangian $\cL^{+4}$ of charge $+4$, which is globally defined on $S^2$
and obeys the analyticity conditions
\begin{align}
D_\alpha^+ \cL^{+4} = \bar D_\dalpha^+ \cL^{+4} = 0~, \qquad
D_\alpha^+ := u_i^+ D_\alpha^i~, \qquad
\bar D_\dalpha^+ := u_i^+ \bar D_\dalpha^i~.
\end{align}
This ensures that $\cL^{+4}$ depends on only half the Grassmann coordinates:
\begin{gather*}
\cL^{+4} = \cL^{+4}(x_A, \q^{+}, \bar \q^{+}, u^\pm)~,\quad
\q^{\pm} := \q^i u_i^\pm~, \quad
\bar\q^{\pm} := \bar \q^i u_i^\pm~, \\
x^m_A := x^m - \ri \,\theta^{+} \sigma^m \bar\theta^{-} - \ri \,\theta^{-} \sigma^m \bar\theta^{+}~.
\end{gather*}
In analogy to $\cN=1$ chiral actions, we can introduce an action as an integral over
half the Grassmann coordinates,\footnote{More precisely, the action involves an
integral over the analytic coordinate $x_A$ rather than $x$. But as with $\cN=1$ chiral
actions, the difference amounts to a total derivative.}
\begin{align}\label{eq_HarmActionPrinc}
S = \int \rd u \int \rd^4x\, \rd^4\theta^+ \cL^{+4}
	= \frac{1}{16} \int \rd^4x \int \rd u\, (D^-)^2 (\bar D^-)^2 \cL^{+4}
\end{align}
where
\begin{align}
D_\alpha^- := u_i^- D_\alpha^i~, \quad
\bar D_\dalpha^- := u_i^- \bar D_\dalpha^i~,
\end{align}
and $\int \rd u$ denotes the integration over $S^2$.

To ensure that the action is real, we require the Lagrangian to be real under
a generalized complex conjugation $\widetilde{\phantom a}$, called smile conjugation.
It corresponds to normal complex conjugation followed by the antipodal map. The smile
conjugates of some relevant quantities are
\begin{align}
\widetilde{u^{i\pm}}= -u_i^\pm~, \qquad
\widetilde{u_i^\pm} = +u^{i\pm}~, \qquad
\widetilde{D^{++}} = D^{++}~.
\end{align}

Within harmonic superspace, the hypermultiplet is described by a complex analytic 
superfield $q^+$ possessing an infinite set of auxiliary fields,
\begin{align}
q^+ = q^i u_i^+ + q^{(ijk)} u_i^+ u_j^+ u_k^- + \cdots
\end{align}
The free hypermultiplet possesses the equation of motion $D^{++} q^+ = 0$, which 
eliminates all but the first term in this expansion, and so on-shell $q^+$
is described by a superfield $q^i$ obeying $D_\alpha^{(i} q^{j)} = \bar D_\dalpha^{(i} q^{j)} = 0$.
More generally, the Lagrangian involving several hypermultiplets $q^{a+}$
is given by \cite{GIOS:HMHS, GIOS:HKgeo} (see also \cite{GIO:Duality})
\begin{align}
\cL^{+4} = \frac{1}{2} q_a^+ D^{++} q^{a+} + H^{+4}~, \qquad H^{+4} = H^{+4}(q^{a+}, u_i^\pm)~,
\end{align}
where $q_a^+ = \Omega_{ab} q^{b+}$ for the canonical symplectic form
$\Omega_{ab}$.\footnote{A more general action is possible, but it can always be put
into this form by a certain gauge transformation.}
The analytic function $H^{+4}$ is a function of $q^{a+}$ and $u_i^\pm$ of
$U(1)$ charge $+4$. In order for the action to be real, we require that
\begin{align}
\widetilde {H^{+4}} = H^{+4}~, \qquad 
\widetilde {q^{a+}} = -q_a^+~, \qquad
\widetilde {q_a^+} = q^{a+}~.
\end{align}
We can choose a basis
\begin{align}
q^{a+} = (q^{I+}, p_I^+)~, \qquad
q_a^+ = (-p_I^+, q^{I+})~, \qquad I=1, \cdots, n
\end{align}
with $\widetilde {q^{I+}} = p_I^+$ and $\widetilde {p_I^+} = -q^{I+}$.
The action becomes in this basis
\begin{align}\label{eq_HarmActionNLS}
S = \int \rd u \int \rd^4x \,\rd^4\q^+ \Big(-p_I^+ D^{++} q^{I+} + H^{+4} \Big)~.
\end{align}
The equations of motion are
\begin{align}\label{eq_HarmEOM}
D^{++} q^{I+} = \frac{\pa H^{+4}}{\pa p_I^+}~, \qquad
D^{++} p_I^+ = -\frac{\pa H^{+4}}{\pa q^{I+}}~.
\end{align}
It was further shown in \cite{GIOS:HKgeo} that \emph{any} $4n$-dimensional
hyperk\"ahler target space geometry can be derived (at least locally) from some
harmonic superspace system of this form.
The target space geometry is encoded in the $\q=0$ parts of
$q^{I+}$ and $p_I^+$, denoted $q^{I+}\vert$ and $p_I^+\vert$,
which turn out to be functions of the $2n$ complex fields
$q^{Ii}(x)$ and $p_{I}^i(x)$. In particular, the
three K\"ahler two-forms $\Omega^{ij}$ describing the hyperk\"ahler geometry
can be encoded in a single two-form\footnote{An equivalent
form of this expression appeared in eq. (A.9) of \cite{GIOS:HKgeo}. Using
their eq. (A.6) and making the canonical choice for the symplectic matrix
yields our eq. \eqref{eq_Harm2Form}.}
\begin{align}\label{eq_Harm2Form}
\Omega^{++} = u_i^+ u_j^+ \Omega^{ij} = (\rd q^{I+} \wedge \rd p_I^+)\vert~,
\end{align}
carrying $U(1)$ charge $+2$. Remarkably, this quantity is formally independent
of $u_i^-$.

It was noted in \cite{GO:Hamiltonian} that the equations \eqref{eq_HarmEOM}
bear a strong resemblance to Hamilton's equations for a symplectic system.
Similarly, \eqref{eq_Harm2Form} resembles the canonical two-form.
This observation is one of the two ingredients of our construction.

Finally, we should mention that although a specific coordinate realization of the harmonics
is usually unnecessary, we will find it useful to introduce a particular parametrization \cite{RoslySchwarz}
in terms of an isotwistor $v^i$ and its conjugate $\bar v_i$ ,
\begin{align}
u^{i+} = \frac{v^i}{\sqrt{v^k \bar v_k}}~, \qquad
u_i^- = \frac{\bar v_i}{\sqrt{v^k \bar v_k}}~.
\end{align}
These relations fix $v^i$ only up to a real factor. Together with
the identification \eqref{eq_HarmU1}, the coordinate $v^i$ is defined
modulo the equivalence relation
\begin{align}\label{eq_ProjEquiv}
v^i \sim c v^i~, \qquad c \in \mathbb C - \{0\}~.
\end{align}
The isotwistor $v^i$ is the homogeneous coordinate for the space $\mathbb CP^1$
and this parametrization realizes the equivalence $\mathbb CP^1 \cong S^2$.
Under smile conjugation, $\widetilde{v^i} = -\veps_{ij} v^j$ and
$\widetilde{\bar v_i} = \veps^{ij} \bar v_j$.

In the northern chart of $\mathbb CP^1$ where $v^\1 \neq 0$, one may introduce
the inhomogeneous coordinate $\z = v^\2 / v^\1$. Then $v^i = v^{\1} (1, \z)$
and the harmonics can be written \cite{GIOS}
\begin{align}\label{eq_HarmZeta}
u^{i+} = \frac{e^{\ri \psi}}{\sqrt{1 + \z \bar \z}} (1, \z)~, \qquad
u_{i}^- = \frac{e^{-\ri \psi}}{\sqrt{1 + \z \bar \z}} (1, \bar\z)~, \qquad
e^{\ri \psi} = \frac{v^\1}{\sqrt{v^\1 \bar v_\1}}~.
\end{align}
The real coordinate $\psi$ parametrizes the $U(1)$ in the coset.
The complex coordinate $\z$ is the usual coordinate describing the Riemann
sphere, arising from stereographic projection from the south pole;
the north pole corresponds to the point $\z=0$ and the south pole to $\z=\infty$.
In these sets of coordinates, the integral over $S^2$ is given by \cite{RoslySchwarz, GIOS}
\begin{align}\label{eq_intS2}
\int \rd u = \frac{1}{2\pi\ri} \int \frac{v^i \rd v_i \wedge \bar v_j \rd \bar v^j}{(v^k \bar v_k)^2}
	= \frac{\ri}{2\pi} \int \frac{\rd \zeta \wedge \rd \bar \zeta}{(1+\z\bar\z)^2}~.
\end{align}

\subsection{Nonlinear sigma models in projective superspace}
Projective superspace \cite{KLR, LR:Projective, LR:SYM} extends the $\cN=2$ supermanifold
$\mathbb R^{4|8}$ by the real projective space $\mathbb CP^1$ parametrized by the complex
isotwistor coordinate $v^i$ defined up to the equivalence \eqref{eq_ProjEquiv}. Because
$\mathbb CP^1 \cong S^2$, it is possible to describe quantities in projective
superspace using the language established for harmonic.\footnote{In fact, both
approaches can be understood as special cases of an isotwistor approach pioneered
by Rosly \cite{Rosly, RoslySchwarz}. See \cite{K:Lectures} for a review.}

Let us begin with the action principle. A projective Lagrangian $\cL^{++}$
is a function of $U(1)$ charge $+2$ which is analytic,
\begin{align}
D_\alpha^+ \cL^{++} = \bar D_\dalpha^+ \cL^{++} = 0~,
\end{align}
real under smile conjugation, $\widetilde{\cL^{++}} = \cL^{++}$,
and constrained to obey
\begin{align}
D^{++} \cL^{++} = 0
\end{align}
in some region on the $S^2$. This implies that the function
$\cL^{(2)} := (v^k \bar v_k) \cL^{++}$ is holomorphic in $v^i$ of degree two,
\begin{align}
\frac{\pa \cL^{(2)}}{\pa \bar v_i} = 0~, \qquad \cL^{(2)}(c v) = c^2 \cL^{(2)}(v)~.
\end{align}
In general, $\cL^{++}$ cannot be globally defined on $S^2$. This is the major
difference with the harmonic action principle: we exchange global definition on $S^2$ for
holomorphy on $\mathbb CP^1$.

Nevertheless, we can construct
an action principle if there exists some contour $C$ in $S^2$ along which $\cL^{++}$
is well-defined:
\begin{align}
S &= \frac{1}{2\pi} \oint_C u^{+} \rd u^+ \int \rd^4x \, (D^-)^4 \cL^{++}~, \qquad
u^{+} \rd u^+ := u^{i+} \rd u_i^+~.
\end{align}
The contour integral is naturally understood as
\begin{align}
\frac{1}{2\pi} \oint_C u^{+} \rd u^+ = \frac{1}{2\pi} \oint_C \frac{v^i \rd v_i}{(v^k \bar v_k)}
	= -\frac{1}{2\pi} \oint_C \frac{ e^{2\ri \psi} \rd \zeta}{1 + \z \bar \z}~.
\end{align}
As in the harmonic case, we may reformulate the action principle to involve
Grassmann integration,
\begin{align}\label{eq_ProjActionPrinc}
S &= \frac{1}{2\pi} \oint_C u^{+} \rd u^+ \int \rd^4x \, \rd^4\q^+\, \cL^{++}~.
\end{align}

General projective hypermultiplets are naturally described by complex superfields $\U^+$
\cite{LR:Projective, GonzalezRey:1} which obey
\begin{align}
D_\alpha^+ \U^+ = \bar D_\dalpha^+ \U^+ = 0~, \qquad D^{++} \U^+ = 0~.
\end{align}
Since $\U^+$ cannot be globally defined, let us restrict to
superfields defined near the north pole, $\zeta=0$. They are given by
a Taylor series
\begin{align}
\U^+ = u^{\1+} \sum_{n=0}^\infty \U_n \zeta^n~.
\end{align}
Such superfields are called ``arctic'' \cite{GonzalezRey:1}. Their smile conjugates
$\breve\U^+ = \widetilde{\U^+}$ are antarctic superfields,
\begin{align}
\breve \U^+ = u^{\2+} \sum_{n=0}^\infty \bar\U_n \left(-\frac{1}{\zeta}\right)^n~,
\end{align}
and are regular at $\zeta=\infty$ in the southern chart of $S^2$.

Nonlinear sigma models are described by projective Lagrangians\footnote{An
equivalent form of this class of Lagrangian appeared originally in \cite{LR:Projective}.
The physical significance of the $u^{i+}$-independent class was first noted
in \cite{GK1, GK2}.}
\begin{align}
\cL^{++} = F^{++} (\U^{I+}, \breve\U_I^+, u^{i+})~, \qquad I=1, \cdots, n
\end{align}
where $F^{++}$ is real under smile-conjugation and possessing $U(1)$ charge $+2$.
By construction, it is analytic and annihilated by $D^{++}$. Introducing the
functions $\U$, $\breve \U$, and $F$ given by\footnote{The original
literature \cite{LR:Properties, GonzalezRey:1}
used $\U$, $\breve \U$ and $F$, which involve only the inhomogeneous coordinate $\zeta$.
When dealing with the superconformal transformation properties of arctic multiplets (see
\cite{K:5Dcomp, K:SPH} where these were first worked out),
the $u^{\1+}$ factor becomes consequential. Although we will not
be dealing explicitly with superconformal models, we find it useful to
keep the $u^{\1+}$.}
\begin{align}
\U^{I+} = u^{\1+} \U^I~, \qquad
\breve\U_I^+ = u^{\2+} \breve\U_I~, \qquad
F^{++} = \ri \,u^{\1+} u^{\2+} F(\U^I, \breve\U_I, \z)~,
\end{align}
the nonlinear sigma model action can be written as an integral over
the $\cN=1$ superspace
\begin{align}
S =  \int \rd^4x\, \rd^2\q_\1 \, \rd^2\bar\q^\1 \,L(\U_n^I, \bar\U_{In})~, \qquad L = \oint \frac{\rd\z}{2\pi \ri \z} F(\U^I, \breve\U_I,\z)~.
\end{align}
The arctic superfield $\U^I$ is made of an infinite number of $\cN=1$ superfields $\U_n^I$,
the lowest two of which, $\Phi^I := \U_0^I$ and $\S^I := \U_1^I$, are constrained,
\begin{align}
\bar D^\ad_\1 \Phi^I = 0~, \qquad (\bar D_\1)^2 \S^I = 0~.
\end{align}
One can eliminate $\U_n^I$ for $n\geq 2$ by solving (formally in general but
explicitly for a broad class of models \cite{GK1, GK2, AN, AKL1,AKL2,KN})
the algebraic equations \cite{GK1, GK2, LR:Properties}
\begin{align}
0 = \frac{\pa L}{\pa \U_n^I} = \oint \frac{\rd \z}{2\pi \ri \z} \frac{\pa F}{\pa \U^I} \z^n ~, \quad n \geq 2
\end{align}
and similarly for $\bar \U_{In}$. The resulting action depends solely on $\Phi^I$
and $\S^I$. The final step is to perform a duality transformation exchanging
the complex linear superfield $\S^I$ for a chiral superfield $\Psi_I$, yielding
an $\cN=1$ nonlinear sigma model action
\begin{align}\label{eq_N1Sigma}
S &= \int \rd^4x\, \rd^2\q_\1 \, \rd^2\bar\q^\1\,K(\Phi^I, \bar\Phi_I, \Psi_I, \bar\Psi^I)~,
\end{align}
where $K$ is the Legendre transform of $L$,
\begin{align}
K(\Phi^I, \bar\Phi_I, \Psi_I, \bar\Psi^I) =
	L(\Phi^I, \bar\Phi_I, \Sigma^I, \bar\Sigma_I) - \Sigma^I \Psi_I - \bar\Sigma_I \bar\Psi^I~.
\end{align}
In accord with the results of \cite{A-GF}, this nonlinear sigma model must be
hyperk\"ahler. One of the three K\"ahler two-forms is manifest, coinciding
with the usual K\"ahler two-form constructed from the K\"ahler potential $K$. 
The structure of the second supersymmetry of the model \eqref{eq_N1Sigma}
establishes that the other two K\"ahler two-forms are given by
$\omega = (\rd \Phi^I \wedge \rd \Psi_I)\vert$ and its complex conjugate \cite{Kuzenko:SigmaDuality, K:CommentsProj}. These three K\"ahler two-forms are related in
the usual way to the three complex structures defining the hyperk\"ahler geometry.

The set of steps described above involving the elimination of auxiliaries
and the Legendre transformation can be succinctly described by the
projective superspace equations of motion \cite{LR:Properties, GK1}
\begin{align}\label{eq_ProjEOM}
\frac{\pa F^{++}}{\pa \U^{I+}} = \ri\, \G_I^{+}~, \qquad
\frac{\pa F^{++}}{\pa \breve\U_I^+} = -\ri\, \breve\G^{I+}~,
\end{align}
where $\G^+$ is required to be an arctic superfield, with
the lowest component in its arctic expansion identified with $\Psi$.
As noted in \cite{HiKLR} (see also the recent discussion in \cite{LR:Properties}),
the three K\"ahler two-forms of this system can be combined
into a single section of a two-form valued $\cO(2)$ bundle,
\begin{align}\label{eq_Proj2FormOld}
\Omega^{++} &= u^{\1+} u^{\2+} \left(
	\frac{1}{\zeta} \rd \Phi^I \wedge \rd \Psi_I
	+ \,\rd \Phi^a \wedge \rd\bar\Phi_b \frac{\pa^2 K}{\pa \Phi^a \pa \bar \Phi_b}
	+ \zeta \rd \bar\Phi_I \wedge \rd \bar\Psi^I \right)\vert~, \eol
\Phi^a &= (\Phi^I, \Psi_I)~, \quad \bar\Phi_a = (\bar\Phi_I, \bar \Psi^I)~.
\end{align}
Using the equations of motion \eqref{eq_ProjEOM}, this can be written in three
equivalent ways \cite{LR:Properties}
\begin{align}\label{eq_Proj2Form}
\Omega^{++} = (\rd \U^{I+} \wedge \rd\G_I^+)\vert
	= -\ri \,\rd \U^{I+} \wedge \rd\breve\U_J^+ \frac{\pa^2 F^{++}}{\pa \U^{I+} \pa\breve\U_J^+} \Big\vert
	= (\rd \breve\U_I^+ \wedge \rd\breve\G^{I+}) \vert
\end{align}
in terms of the full arctic and antarctic expansions.
One may interpret the on-shell Lagrangian $F^{++}$ as a generating function for the
canonical transformation connecting
the ``arctic coordinates'' $\U^{I+}$ and $\G_I^+$ to the ``antarctic coordinates''
$\breve\U_I^+$ and $\breve\G^{I+}$, while preserving the symplectic two-form
$\Omega^{++}$ \cite{LR:Properties}.
These observations regarding $F^{++}$ form the second ingredient to the construction
we will present in this paper. For now, let us merely point out the striking similarity
between \eqref{eq_Harm2Form} and \eqref{eq_Proj2Form}.

\subsection{Hamiltonian mechanics and canonical transformations}
The main construction of our paper is to use a complexified harmonic superspace
to map the harmonic description of a nonlinear
sigma model to a projective description. To keep our
presentation self-contained, we briefly review an alternative
construction of the action principle in Hamiltonian mechanics that we
will encounter shortly in a more complicated setting.

Recall that in the usual formulation of one-dimensional Hamiltonian
mechanics,\footnote{The generalization to
$n$ dimensions is straightforwardly accomplished by replacing $q \rightarrow q^I$
and $p \rightarrow p_I$.}
one introduces the action functional $S[q(t); t_0, t_1]$ defined by
\begin{align}
S = \int_{t_0}^{t_1} \rd t\, L = \int_{t_0}^{t_1} \rd t\, (p \dot q - H )~.
\end{align}
The initial and final times $t_0$ and $t_1$ are fixed, as are the trajectory
endpoints $q_0 = q(t_0)$ and $q_1 = q(t_1)$.
Taking the path $q(t)$ which extremizes the action, $S$ becomes a function
of $q_0$, $q_1$, $t_0$, and $t_1$, with $q(t)$ required to obey Hamilton's equations
\begin{align}\label{eq_Hamilton}
\frac{\rd q}{\rd t} = \frac{\pa H}{\pa p}~, \qquad
\frac{\rd p}{\rd t} = -\frac{\pa H}{\pa q}~.
\end{align}
Solutions to Hamilton's equations automatically preserve the canonical two-form
\begin{align}
\omega = \rd q \wedge \rd p~,
\end{align}
so that $\omega$ is independent of $t$.
The partial derivatives of $S$ with respect to $q_0$ and $q_1$ are given by
\begin{align}\label{eq_paS}
\frac{\pa S}{\pa q_1} = p_1~, \qquad
\frac{\pa S}{\pa q_0} = -p_0~,
\end{align}
which identifies the action $S$ as the generating function (of the first type)
for the canonical transformation associated with time evolution from $t_0$ to $t_1$.

We can construct another generating function (of the second type)
for this canonical transformation via a Legendre transformation:
\begin{align}
F(q_0, p_1, t_0, t_1) := p_1 q_1 - S~,
\end{align}
which obeys
\begin{align}\label{eq_paF}
\frac{\pa F}{\pa q_0} = p_0~, \qquad
\frac{\pa F}{\pa p_1} = q_1~.
\end{align}
The function $F$ can equivalently be written
\begin{align}\label{eq_defF}
F = \frac{1}{2} p_0 q_0 + \frac{1}{2} p_1 q_1
+ \int_{t_0}^{t_1} \rd t \,\Big(
	H - \frac{1}{2} p \dot q + \frac{1}{2} q \dot p
\Big)~,
\end{align}
which places $q$ and $p$ on a more symmetric footing.
This alternative action is just as fundamental as $S$. Let us now take
\eqref{eq_defF} as the \emph{definition} for $F$
with $q(t)$ and $p(t)$ arbitrary paths constrained only by
$q(t_0) = q_0$ and $p(t_1) = p_1$. Under arbitrary deformations of
the paths that leave $q_0$ and $p_1$ invariant, $F$ is extremized
precisely for those solutions $q(t)$ and $p(t)$ which obey Hamilton's
equations.
This alternative action for Hamiltonian mechanics works even for the case
of a vanishing Hamiltonian where the usual action principle fails. For such a
trivial system, Hamilton's equations are solved by $q(t) = q_0 = q_1$ and
$p(t) = p_0 = p_1$. While the usual action $S$ vanishes in such a case,
the symplectic action $F$ is given by $F = q_0 p_1$ and
yields the correct canonical transformation.

It is possible to draw an analogy between harmonic and projective
descriptions of nonlinear sigma models and the symplectic system described
above. The harmonic equations of motion \eqref{eq_HarmEOM} are analogous to
Hamilton's equations \eqref{eq_Hamilton}, as noted in \cite{GO:Hamiltonian}.
The projective Lagrangian $F^{++}$ is analogous to the action $F$,
as both act (on-shell) as the generating function for a canonical
transformation on their respective symplectic systems.

The remainder of this paper is devoted to fleshing out this analogy. In section \ref{sect_2},
we show how to perform a certain contour deformation so that we can
interpret the equations \eqref{eq_HarmEOM} in terms of a one-parameter Hamiltonian system.
The actual construction is given in section \ref{sect_3}, where we explain
how it works with the trivial case of a free hypermultiplet and then
test it with the cases of Taub-NUT and Eguchi-Hanson.
We conclude with a brief discussion.

\section{An alternative coordinate system and the complexified $S^2$}\label{sect_2}
We begin with the harmonics in the form \eqref{eq_HarmZeta}, with
$\z$ describing the inhomogeneous coordinate on $\mathbb CP^1$.
Following \cite{JS1}, we introduce a real coordinate $t$ given
by\footnote{Our conventions differ slightly from \cite{JS1} by the replacement
$t \rightarrow 1-t$.}
\begin{align}\label{eq_deft}
t = \frac{\z \bar\z}{1+\z\bar\z}~.
\end{align}
The north pole lies at $t=0$ and the south pole lies at $t=1$.
The harmonics may now be parametrized in terms of $u^{\1+}$, $\zeta$, and $t$:
\begin{align}\label{eq_HarmComplex}
u^{i +} &= (u^{\1+}, u^{\2+}) = u^{\1+} (1, \z)~, \eol
u_i^- &= \frac{1}{u^{\1+}} (1-t,0) + \frac{1}{u^{\2+}} (0,t) = \frac{1}{u^{\1+}} \Big(1-t, \frac{t}{\z}\Big)~.
\end{align}
The $S^2$ is described by $\z$ and $t$; the complex variable $u^{\1+}$ will drop
out of all explicit calculations.
Under smile conjugation,
\begin{align}\label{eq_TildeC}
\widetilde{u^{\1+}} =  \z u^{\1+}~, \qquad
\widetilde{\z} = -\frac{1}{\zeta}~, \qquad
\widetilde{t} \rightarrow 1-t~.
\end{align}
The advantage of this coordinate system is that the operator $D^{++}$ takes an especially simple form:
\begin{align}
D^{++} = u^{\1+} u^{\2+} \frac{\pa}{\pa t}~.
\end{align}
If we could interpret the real variable $t$ as an independent coordinate
(\textit{i.e.} relax the constraint \eqref{eq_deft}), then the harmonic equations
of motion \eqref{eq_HarmEOM} would resemble Hamilton's equations.

In this coordinate system, the harmonic measure becomes
\begin{align}
\int \rd u = \int_0^{1} \rd t \oint_{C(t)} \frac{\rd \zeta}{2\pi \ri \z}
\end{align}
where $C(t)$ is a contour along the latitude of fixed $|\z|^2 = t / (1-t)$.
Because harmonic actions are required to be globally defined, the only
harmonic integrals one encounters are always of the form
$\int \rd u \, F$
where $F(u)$ is a globally defined weight-zero function with the expansion
\begin{align}\label{eq_Fweight0}
F(u) = F(\z,t) = f + f^{(ij)} u_i^+ u_j^- + \cdots~.
\end{align}
Since the harmonic integral selects out only the zero mode, $\int \rd u \,F = f$,
we are free to deform the $\zeta$ contour to be any latitude we wish
and we maintain this property. (Equivalently, we treat $F(\z,t)$ as
holomorphic in $\z$ with singularities only at $\z=0$ and $\z=\infty$.)
This means we can take
\begin{align}
\int \rd u \,F(u) = \int_0^1 \rd t \oint_{C(t)} \frac{\rd \zeta}{2\pi \ri \z} \,F(\z,t)
	\Longrightarrow  \oint_C \frac{\rd \zeta}{2\pi \ri \z} \int_0^1 \rd t \,F(\z,t)
\end{align}
where $\z$ is no longer constrained to obey \eqref{eq_deft} and $C$
is now a $t$-independent contour. (A nearly identical approach was used in \cite{JS1}.)
For definiteness, we will take $C$ to be along the equator.
For a function $F(\z,t)$ of the form \eqref{eq_Fweight0}, this integral again selects
out the zero mode.

This approach of deforming the $\zeta$ contour can be interpreted as performing
a certain complexification of the $S^2$, a concept which has reappeared several times
in the harmonic superspace literature.\footnote{See \textit{e.g.} \cite{GIOS:Conformal, GIOS}
for the relevance of complex harmonics for superconformal transformations,
and \cite{GIO:Quat} for their importance in describing quaternionic sigma models.}
A complete discussion is beyond the scope of this paper, but we briefly discuss some
details relevant to our construction in appendix \ref{app_cS2}.

Now let us apply this alternative coordinate system to a harmonic superspace action.
Given the action \eqref{eq_HarmActionPrinc}
\begin{align}
S = \frac{1}{16} \int \rd^4x \int \rd u \,(D^-)^2 (\bar D^-)^2 \cL^{+4}~,
\end{align}
we may rearrange the integrand and measure to the form
\begin{align}
S = \frac{1}{16} \int \rd^4x \oint_C \frac{\rd \zeta}{2\pi \ri \zeta}
	\frac{1}{(u^{\1+} u^{\2+})^2} \int_0^1 \rd t \,(D^\1)^2 (\bar D_\1)^2 \cL^{+4}~,
\end{align}
after discarding a total derivative. Now define the projective Lagrangian
\begin{align}\label{eq_ProjLFromHarmL}
\cL^{++} = \frac{\ri}{u^{\1+} u^{\2+}} \int_0^1 \rd t\,\cL^{+4}~.
\end{align}
It is analytic by construction. After the complexification,
$\cL^{+4}$ is a function of the auxiliary coordinates $u^{\1+}$, $\z$, and $t$,
and so $\cL^{++}$ is a function of only $u^{\1+}$ and $\z$.
The corresponding action can be written
\begin{align}
S = \frac{1}{2\pi} \oint_C u^{+} \rd u^+ \int \rd^4x \,\rd^4\q^+\, \cL^{++}~,
\end{align}
which is exactly the form of a projective superspace action.\footnote{A similar procedure was
used in \cite {K:DP} by replacing regular functions on $S^2$ with functions which
were holomorphic everywhere except for a small region near the poles.}

We would like to check if \eqref{eq_ProjLFromHarmL} generates a reasonable projective action
for a simple case.
Note that the expression \eqref{eq_ProjLFromHarmL} is singular in two ways. First, it involves
the harmonic prefactor $1 / u^{\1+} u^{\2+}$ which diverges at
the poles. Second, the complexification of the harmonics will lead
well-defined functions on $S^2$ to develop singularities (as noted in \cite{JS1}).
The second issue is a more complicated one, so for the moment
let's stick to an action involving multiplets which are always globally defined
even in the projective setting. The simplest such multiplet is 
the tensor multiplet $G^{++} = G^{ij} u_i^+ u_j^+$, which is a globally defined
$\cO(2)$ multiplet in both the harmonic and projective descriptions.
It possesses a unique superconformal action, the improved tensor multiplet action,
which takes a quite different form depending on whether one uses a harmonic or
a projective realization. In the projective setting, the Lagrangian can be
written \cite{KLR}
\begin{align}\label{eq_ImpTensProj}
\cL^{++} = -G^{++} \log (G^{++} / \ri \,u^{\1+} u^{\2+})
\end{align}
where its resemblance to the $\cN=1$ analogue is striking.
In the harmonic realization, the action is quite different \cite{GIO:Tensor} (see also \cite{GIOS}).
One requires the introduction of an auxiliary isovector $c^{++} = c^{ij} u_i^+ u_j^+$,
normalized so that $c^2 = \frac{1}{2} c^{ij} c_{ij} = 1$,
in terms of which the Lagrangian can be written
\begin{align}\label{eq_ImpTensHarm}
\cL^{+4} = \left(\frac{\ell^{++}}{1+\sqrt{1+\ell^{++} c^{--}}}\right)^2~, \qquad
\ell^{++} = G^{++} - c^{++}~.
\end{align}
A global $SU(2)$ rotation allows the choice $c_{\1\2} = \ri$ so that
$c^{++} = 2 \ri \,u^{\1+} u^{\2+}$.
Now we complexify the harmonics \eqref{eq_HarmComplex} so that
\begin{align}
c^{--} = -2 \ri \,u_{\1}^- u_{\2}^- = \frac{2}{\ri \,u^{\1+} u^{\2+}} \, t (1-t)~.
\end{align}
Performing the $t$ integral in \eqref{eq_ProjLFromHarmL} immediately gives
\begin{align}\label{eq_ImpTensDerived}
\cL^{++} = G^{++} - c^{++} - G^{++} \log (G^{++} / c^{++})~,
\end{align}
which agrees with \eqref{eq_ImpTensProj} up to terms which do not contribute to the action.
This verifies the reasonableness of the complexification procedure for
a fairly nontrivial example. Now let us move on to a more complicated multiplet.

\section{Deriving projective nonlinear sigma models and solutions from harmonic superspace}\label{sect_3}
In this section, we first give our main result: the mapping of general nonlinear sigma models
in harmonic superspace to nonlinear sigma models in projective superspace.
For notational simplicity, we restrict to the case of a single hypermultiplet,
but our results generalize straightforwardly to the case of $n$ hypermultiplets.
We also give the explicit on-shell relation for three example cases: the free
hypermultiplet, Taub-NUT, and Eguchi-Hanson.

\subsection{Nonlinear sigma models on the complexified $S^2$}\label{sect_3.1}
Following \cite{JS1}, we begin by complexifying the functions $q^+$ and $p^+$,
\begin{align}
q^+ \rightarrow Q^+ = Q^+(u^{\1+}, \z, t)~, \qquad
p^+ \rightarrow P^+ = P^+(u^{\1+}, \z, t)~,
\end{align}
where $\zeta$ is no longer constrained to
obey \eqref{eq_deft}.\footnote{The
$u^{\1+}$ dependence of these expressions is trivial: $Q^+(u^{\1+}, \z, t) = u^{\1+} Q(\z,t)$
and similarly for $P^+$.}
The analytic constraints remain
\begin{align}
D_\alpha^+ Q^+ = \bar D_\dalpha^+ Q^+ = 0~, \qquad
D_\alpha^+ P^+ = \bar D_\dalpha^+ P^+ = 0~,
\end{align}
and so $Q^+$ and $P^+$ are analytic superfields.
We introduce obvious notation for the boundary values of $Q^+$ and $P^+$ at $t=0$ and $t=1$, respectively,
\begin{subequations}
\begin{align}
Q_0^+ &:= Q^+(u^{\1+}, \z, 0)~, \qquad Q_1^+ := Q^+(u^{\1+}, \z, 1) \\
P_0^+ &:= P^+(u^{\1+}, \z, 0)~, \qquad P_1^+ := P^+(u^{\1+}, \z, 1)~.
\end{align}
\end{subequations}
If $q^+$ and $p^+$ are globally defined functions on $S^2$, then $Q_0^+$
and $P_0^+$ must be regular at $\z=0$, and $Q_1^+$ and $P_1^+$
must be regular at $\z = \infty$. We identify
\begin{align}
Q_0^+ \equiv \U^+~, \qquad P_0^+ \equiv \G^+~, \qquad
Q_1^+ \equiv -\breve \G^+~, \qquad P_1^+ \equiv \breve \U^+
\end{align}
where $\U^+$ and $\G^+$ are arctic multiplets and $\breve\U^+$ and $\breve \G^+$
are antarctic.
Note that the smile conjugation of harmonic superspace \eqref{eq_TildeC}
includes a reflection in the real parameter $t$.
The above observations are in accord with the results of \cite{JS1}.

We want to construct an action for $Q^+$ and $P^+$ so that they end up obeying
the complexified version of the harmonic equations \eqref{eq_HarmEOM},
\begin{align}\label{eq_HarmEOMComplex}
D^{++} Q^+ = \frac{\pa H^{+4}}{\pa P^+}~, \qquad
D^{++} P^+ = - \frac{\pa H^{+4}}{\pa Q^+}~.
\end{align}
Because $t$ and $\z$ are now independent variables, these equations
are simply Hamilton's equations for a symplectic system: the
``position'' $Q^+$ and ``momentum'' $P^+$
merely possess additional dependence on the coordinates $\zeta$ and $u^{\1+}$.

To do this, it turns out that we need to relax the assumption that $Q_1^+$ is regular at $\z=\infty$
and $P_0^+$ is regular at $\z=0$. In other words, we assume $\G^+$ and $\breve \G^+$
to possess full Laurent expansions.\footnote{The convergence of these expansions is
unclear to us. It seems reasonable that $\G^+$ and $\breve \G^+$ should be
taken initially to be ``tropical'' superfields in the language of projective superspace,
with singularities allowed at both $\z=0$ and $\z=\infty$.}
However, we retain the property that
$Q_0^+ \equiv \U^+$ is regular at $\z=0$ and similarly,
$P_1^+ \equiv \breve\U^+$ remains regular at $\z=\infty$. (This is in accord
with the analogous symplectic problem, where we fix $q_0$ and $p_1$
but do not fix the opposite endpoints.)
Then, inspired by eq. \eqref{eq_defF}, we postulate the following action on the
complexified $S^2$:
\begin{align}\label{eq_cHarmActionNLS}
S &= \int_{\cP^2} \rd u\, \int \rd^4x \,\rd^4\q^+ \Big(-\frac{1}{2} P^+ D^{++} Q^{+} + \frac{1}{2} Q^{+} D^{++} P^+ + H^{+4} \Big)
	\eol & \quad
	+ \frac{\ri}{4\pi} \oint_C u^+ \rd u^+ \int \rd^4x \,\rd^4\q^+ Q_0^+ P_0^+
	+ \frac{\ri}{4\pi} \oint_C u^+ \rd u^+ \int \rd^4x \,\rd^4\q^+ Q_1^+ P_1^+~.
\end{align}
The first integral is an obvious generalization of the harmonic action and involves
an integral over a region $\cP^2$ of the complexified $S^2$ (see the discussion
in appendix \ref{app_cS2}):
\begin{align}
\int_{\cP^2} \rd u := \oint_C \frac{\rd \zeta}{2\pi \ri \z} \int_0^1 \rd t~.
\end{align}
We have taken the contour $C$ to be along the equator with $\abs{\zeta} = 1$.
The second and third integrals resemble ``surface terms'' and involve the same contour $C$.
In this form, the action bears a very close resemblance to the harmonic superspace
action \eqref{eq_HarmActionNLS}. In fact, if we required $Q^+$ and $P^+$ to be globally
defined on the real $S^2$, then $P_0^+ \equiv \G^+$ would be arctic
(and $Q_1^+ \equiv -\breve\G^+$ would be antarctic) and the
``surface terms'' wouild vanish, giving back the original harmonic action.

Now let us rewrite the action in a more suggestive form.
Noting that
\begin{align}
\frac{1}{2\pi} \oint_C u^+ \rd u^+ = -\oint_C \frac{\rd \zeta}{2\pi \ri \zeta}\, \ri \,u^{\1+} u^{\2+}~,
\end{align}
we can write
\begin{align}
S &= \frac{1}{2\pi} \oint_C u^+ \rd u^+ \int \rd^4x \,\rd^4\q^+ F^{++}~,
\end{align}
where the analytic function $F^{++}$ is given by
\begin{align}\label{eq_dpcNLS}
F^{++} &=
	\frac{\ri}{2} Q_0^+ P_0^+
	+ \frac{\ri}{2} Q_1^+ P_1^+
	+ \frac{\ri}{u^{\1+} u^{\2+}} \int_0^1 \rd t\,
	\Big(H^{+4} -\frac{1}{2} P^+ D^{++} Q^{+} + \frac{1}{2} Q^{+} D^{++} P^+ \Big)~.
\end{align}
Note that $F^{++}$ is holomorphic in $\zeta$ due to the complexification and
satisfies all the requirements of a projective Lagrangian. It is also completely
analogous to the symplectic action \eqref{eq_defF} discussed in the introduction.

Let us vary the Lagrangian. We find
\begin{align}\label{eq_varyF}
\delta F^{++} &= \ri \,\delta Q_0^+ P_0^+ + \ri \,Q_1^+ \delta P_1^+
	\eol & \quad
	+ \frac{\ri}{u^{\1+} u^{\2+}} \int_0^1 \rd t\, \bigg\{
	\delta Q^+ \Big(D^{++} P^+ + \frac{\pa H^{+4}}{\pa Q^+}\Big)
	- \delta P^+ \Big(D^{++} Q^+ - \frac{\pa H^{+4}}{\pa Q^+}\Big) \bigg\}~.
\end{align}
If we fix the endpoints $Q_0^+ \equiv \U^+$ and $P_1^+ \equiv \breve\U^+$,
we recover the complexified version of the harmonic equations \eqref{eq_HarmEOMComplex}.
In principle, these may be solved in terms of $\U^+$ and $\breve\U^+$.
Reinserting the solution into the Lagrangian, we find
\begin{align}
F^{++} = F^{++}(\U^+, \breve\U^+, u^+)~.
\end{align}
This exactly yields the projective superspace action principle.

Now suppose we allow the endpoints to vary. We find that in order for
the action to vanish, we must have
\begin{align}
0 &= \frac{1}{2\pi} \oint_C u^+ \rd u^+ \int \rd^4x \,\rd^4\q^+
	\Big(\ri \,\delta Q_0^+ P_0^+ + \ri \,Q_1^+ \delta P_1^+\Big)
\end{align}
where $\delta Q_0^+ \equiv \delta \U^+$ is arctic and
$\delta P_1^+ \equiv \delta \breve \U^+$ is antarctic.
The solution is familiar from projective superspace:
$P_0^+ \equiv \Gamma^+$ must be arctic and similarly
$Q_1^+ \equiv -\breve\Gamma^+$ must be antarctic.
We also discover
\begin{align}
\frac{\pa F^{++}}{\pa \U^+} = \ri \,\Gamma^+~, \qquad
\frac{\pa F^{++}}{\pa \breve \U^+} = -\ri \,\breve\Gamma^+~,
\end{align}
when $F^{++}$ is written in terms of the endpoints
$\U^+$ and $\breve\U^+$.

Remarkably, the same action \eqref{eq_cHarmActionNLS} yields
the harmonic superspace action when restricted to globally
defined functions on the real $S^2$ and also the projective
superspace action when put partly on-shell.
Similarly, the same solutions, the trajectories $Q^+$ and $P^+$,
must reduce on the real $S^2$ to the harmonic solutions $q^+$
and $p^+$ and  at their endpoints to the projective solutions
$\U^+$ and $\Gamma^+$.
Moreover, the harmonic and projective descriptions of the
two-form $\Omega^{++}$, eqs. \eqref{eq_Harm2Form} and \eqref{eq_Proj2Form},
become transparently identified.

Let us see how this construction works in practice.

\subsection{Examples}
We will restrict our examples to on-shell multiplets (\textit{i.e.} globally
defined multiplets on the real $S^2$) because their
solutions in harmonic superspace are available.
Our first example is a very simple one: the free hypermultiplet.
This was considered already in \cite{K:DP, JS1},
but we include it here for completeness.
Then we address the more challenging cases of Taub-NUT and Eguchi-Hanson.

\subsubsection*{Free hypermultiplet}
The free hypermultiplet corresponds to the harmonic action \eqref{eq_HarmActionNLS}
with a vanishing Hamiltonian $H^{+4} = 0$. The equations of motion are easily
solved by
\begin{align}
q^+ = q^i u_i^+~, \qquad p^+ = p^i u_i^+~.
\end{align}
Now we consider its complexification,
\begin{align}
Q^+ = q^i u_i^+~, \qquad P^+ = p^i u_i^+~.
\end{align}
Since these have no dependence on $u_i^-$, they are $t$-independent:
\begin{align}
Q^+ = Q_0^+ = Q_1^+ = \U^+ = -\breve \Gamma^+~, \qquad
P^+ = P_0^+ = P_1^+ = \breve \U^+ = \Gamma^+~.
\end{align}
The projective Lagrangian $F^{++}$ is easy to calculate:
\begin{align}
F^{++} =
	\frac{\ri}{2} Q_0^+ P_0^+
	+ \frac{\ri}{2} Q_1^+ P_1^+ = \ri \U^+ \breve \U^+~.
\end{align}
The solution clearly obeys the projective superspace equations of motion \eqref{eq_ProjEOM}.

\subsubsection*{Taub-NUT}
Next we consider the nontrivial case of Taub-NUT.
The harmonic action possesses the Hamiltonian
\begin{align}
H^{+4} = -\frac{\lambda}{2} (q^+ p^+)^2~,
\end{align}
where $\lambda$ is a real parameter. The solution is known \cite{GIOS:HMHS}:
\begin{align}
q^+ &= f^+ \exp \left(-\frac{\lambda}{2} (f^+ \tilde f^- + f^- \tilde f^+) \right)~, \quad
p^+ = \tilde f^+ \exp \left(+\frac{\lambda}{2} (f^+ \tilde f^- + f^- \tilde f^+) \right)~, \eol
f^\pm &:= f^i u_i^\pm~, \qquad
\tilde f^\pm := \bar f^i u_i^{\pm}~,
\end{align}
for some superfield $f_i$ and its complex conjugate $\bar f^i$. Note that
$q^+ p^+ = f^+ \tilde f^+$ is independent of $u_i^-$, and the Hamiltonian
\begin{align}
H^{+4} = - \frac{\lambda}{2} (f^+ \tilde f^+)^2~
\end{align}
is also independent of $u_i^-$.

Now let us complexify. We take
\begin{align}
Q^+ = f^+ \exp \left(-\frac{\lambda}{2} (f^+ \tilde f^- + f^- \tilde f^+) \right)~,\quad
P^+ = \tilde f^+ \exp \left(+\frac{\lambda}{2} (f^+ \tilde f^- + f^- \tilde f^+) \right)
\end{align}
where $u^{i+}$ and $u_i^-$ are given by \eqref{eq_HarmComplex} but without imposing
\eqref{eq_deft}. We observe that
$Q^+ P^+ = Q_0^+ P_0^+ = Q_1^+ P_1^+$
is independent of $u_i^-$ and therefore of $t$. The on-shell projective Lagrangian is given by
\begin{align}
F^{++} &=
	\frac{\ri}{2} Q_0^+ P_0^+
	+ \frac{\ri}{2} Q_1^+ P_1^+
	+ \frac{\ri}{u^{\1+} u^{\2+}} \int_0^1 \rd t\,
	\Big(H^{+4} -\frac{1}{2} P^+ D^{++} Q^{+} + \frac{1}{2} Q^{+} D^{++} P^+ \Big) \eol
	&= 
	\ri \,Q_0^+ P_0^+
	+ \frac{\ri}{u^{\1+} u^{\2+}} \int_0^1 \rd t\,
	\Big(H^{+4} -\frac{1}{2} P^+ \frac{\pa H^{+4}}{\pa P^+}
	- \frac{1}{2} Q^{+} \frac{\pa H^{+4}}{\pa Q^+} \Big) \eol
	&=  \ri \,Q_0^+ P_0^+ - \frac{\ri}{u^{\1+} u^{\2+}} H^{+4}~.
\end{align}
We can write this in terms of $f^+$ and $\tilde f^+$ as
\begin{align}
F^{++} &= \ri f^+ \tilde f^+ - \frac{\lambda}{2\ri \,u^{\1+} u^{\2+}} (f^+ \tilde f^+)^2~.
\end{align}
For shorthand, let us introduce the real analytic multiplet
\begin{align}
U^{++} = \ri \,Q^+ P^+~.
\end{align}
On-shell, we see that $U^{++}$ is actually an $\cO(2)$ multiplet,
\begin{align}
U^{++} = \ri f^+ \tilde f^+ \quad\implies\quad D^{++} U^{++} = 0~.
\end{align}
We also introduce the constant $c^{++} = \ri \,u^{\1+} u^{\2+}$
for convenience. Then we have
\begin{align}\label{eq_TN_G}
F^{++} = U^{++} + \frac{\lambda}{2} \frac{(U^{++})^2}{c^{++}}~.
\end{align}
This is not useful quite yet since we require $F^{++}$ to be written
in terms of $\U^+$ and $\breve\U^+$. Observing that
\begin{subequations}\label{eq_TNsolution}
\begin{align}
\U^+ &\equiv Q_0^+ = f^+ \exp \left(-\frac{\lambda}{2 u^{\1+}} (f^+ \bar f^{\1} + f^{\1} \tilde f^+) \right)~, \\
\breve \U^+ &\equiv P_1^+ = \tilde f^+ \exp \left(+\frac{\lambda}{2 u^{\2+}} (f^+ \bar f^{\2} + f^{\2} \tilde f^+) \right)~,
\end{align}
\end{subequations}
we find
\begin{align}\label{eq_TN_redef}
\ri \,\U^+ \breve \U^+ &= \ri f^+ \tilde f^+ \exp \left(\lambda \,\frac{f^+ \tilde f^+}{u^{\1+} u^{\2+}}\right)
	= U^{++} \exp \left(\lambda \,U^{++} / c^{++}\right)~.
\end{align}
This relation can be inverted to solve for $U^{++}$ in terms of
$\U^+ \breve \U^+$:
\begin{align}
U^{++} = \frac{c^{++}}{\lambda} W\Big(\ri \lambda \U^+ \breve \U^+ / c^{++}\Big)~,
\end{align}
where $W(z)$ is the Lambert $W$-function.\footnote{The
Lambert $W$-function $W(z)$ is defined as the solution
$w = W(z)$ to the equation $z = w e^w$.}
Reinserting into the Lagrangian gives
\begin{align}\label{eq_TNactionDerived}
F^{++} = \frac{c^{++}}{\lambda} W\Big(\ri \lambda \U^+ \breve \U^+ / c^{++}\Big)
	+ \frac{c^{++}}{2 \lambda} W\Big(\ri \lambda \U^+ \breve \U^+ / c^{++}\Big)^2~.
\end{align}

Now let us check that this Lagrangian is indeed the Taub-NUT Lagrangian in
projective superspace and that \eqref{eq_TNsolution} gives the solution.
The Taub-NUT action is expressed naturally as the sum of the free and improved tensor
multiplet actions \cite{HiKLR}
\begin{align}
\cL^{++} = G^{++} - G^{++} \log\left(G^{++} / \ri \U^+ \breve \U^+ \right) - \frac{\lambda}{2} \frac{(G^{++})^2}{c^{++}}~,
\end{align}
where $G^{++}$ is an $\cO(2)$ (or tensor) multiplet and $c^{++} = \ri \,u^{\1+} u^{\2+}$
is a real constant.\footnote{We have included the $G^{++}$ term for convenience.
It does not contribute to the action.}
The arctic multiplet $\U^+$ is a pure gauge degree of freedom so long as $G^{++}$
is an $\cO(2)$ multiplet. We may derive this from the first-order Lagrangian
\begin{align}
\cL^{++}_{\rm F.O.} = U^{++} - U^{++} \log\left(U^{++} / \ri \U^+ \breve \U^+ \right) - \frac{\lambda}{2} \frac{(U^{++})^2}{c^{++}}~,
\end{align}
where $U^{++}$ is a real unconstrained analytic multiplet. Now the arctic
multiplet's presence is consequential: applying
the $\U^+$ equation of motion leads to the condition that $U^{++}$
is an $\cO(2)$ multiplet. Instead, let us apply the $U^{++}$ equation of motion.
This leads to
\begin{align}\label{eq_Usoln}
\ri \U^+ \breve \U^+ = U^{++} \exp(\lambda U^{++} / c^{++}) \quad\implies\quad
U^{++} = \frac{c^{++}}{\lambda} W\Big(\ri \lambda \U^+ \breve \U^+ / c^{++}\Big)~.
\end{align}
Reinserting this into the Lagrangian leads to the hypermultiplet action for Taub-NUT:
\begin{align}
F^{++} = \frac{c^{++}}{\lambda} W\Big(\ri \lambda \U^+ \breve \U^+ / c^{++}\Big)
	+ \frac{c^{++}}{2 \lambda} W\Big(\ri \lambda \U^+ \breve \U^+ / c^{++}\Big)^2~.
\end{align}
This indeed matches \eqref{eq_TNactionDerived}.
Now let's demonstrate that \eqref{eq_TNsolution} gives the correct solution.
Recall that $\U^+$ must obey the equation of motion
\begin{align}
\frac{\pa F^{++}}{\pa \U^+} = \ri \,\G^+
\end{align}
where $\G^+$ is some arctic multiplet. This can be written
\begin{align}
\ri \,\G^+ = \frac{\pa \cL^{++}_{\rm F.O.}}{\pa U^{++}} \frac{\pa U^{++}}{\pa \U^+}
	= \Big(1+\lambda U^{++} / c^{++} \Big) \frac{\pa U^{++}}{\pa \U^+} 
\end{align}
where $U^{++}$ satisfies \eqref{eq_Usoln}. Imposing the solution \eqref{eq_TNsolution}, we see
that \eqref{eq_Usoln} is solved by $U^{++} = \ri \,f^+ \tilde f^+$. Taking the $\U^+$ derivative
of \eqref{eq_Usoln}, we find
\begin{align}
\ri\, \breve \U^+ = \Big(1+\lambda U^{++} / c^{++} \Big) \exp\left(\lambda U^{++} / c^{++}\right)
	\frac{\pa U^{++}}{\pa\U^+}~,
\end{align}
which implies that
\begin{align}
\G^+ = \exp\left(-\lambda U^{++} / c^{++}\right) \breve\U^+
	= \tilde f^+ \exp\left(\frac{\lambda}{2 u^{\1+}}(f^+ \bar f^\1 + f^\1 \tilde f^+) \right)~.
\end{align}
As required, this expression is arctic. As a check, we note that $\G^+$ is indeed given by
the $t=0$ boundary of the complexified $P^+$ hypermultiplet: $\G^+ = P^+(u^{\1+}, \z, 0)$.
Similarly, $\breve \Gamma^+$ is antarctic and given by $\breve \G^+ = - Q^+(u^{\1+}, \z, 1)$.
Thus \eqref{eq_TNsolution} gives the solution for Taub-NUT in projective
superspace.

It should be mentioned that although \eqref{eq_TNsolution} is the solution,
it is not in a useful form because the superfields $f_i$ and $\bar f^i$ are
quite complicated. One should introduce the $\cN=1$ chiral superfields $\Phi$
and $\Psi$, which are found by taking the $\z=0$ limits of $\U^+$
and $\G^+$, respectively. They are
\begin{align}
\Phi = - f_\1 \exp\Big(\frac{\lambda}{2} (f_\1 \bar f^\1 - f_\2 \bar f^\2) \Big)~, \qquad
\Psi = \bar f^\2 \exp\Big(-\frac{\lambda}{2} (f_\1 \bar f^\1 - f_\2 \bar f^\2) \Big)~.
\end{align}
In principle, one can solve for $f_\1$ and $f_\2$ in terms of $\Phi$ and $\Psi$
and then reinsert the result into the expressions for $\U^+$ and $\G^+$.
One can even (in principle) construct the $\cN=1$ Lagrangian $K$ indirectly
using the explicit forms of $\Omega^{++}$, \eqref{eq_Proj2FormOld} and \eqref{eq_Proj2Form},
to extract the K\"ahler metric and then to integrate it.

\subsubsection*{Eguchi-Hanson}
Our last example is Eguchi-Hanson. The harmonic action involves
the Hamiltonian \cite{Gibbons:1987pk} (see also \cite{GIOT})
\begin{align}\label{eq_EHHarm}
H^{+4} = \frac{2}{r^2} (g^{++})^2~, \qquad
g^{++} := \frac{\ell^{++}}{1+\sqrt{1+4 \ell^{++} \xi^{--} / r^4}}~, \qquad \ell^{++} := -\ri \,q^+ p^+~,
\end{align}
where $\xi^{--} = \xi^{ij} u_i^- u_j^-$ is a constant isotriplet
normalized as $r^2 = \sqrt{2 \xi^{ij} \xi_{ij}}$. We specialize to the case
$\xi_{\1\2} = \ri \, r^2 / 2$ and $\xi_{\1\1} = \xi_{\2\2} = 0$.
One solution is
\begin{subequations}\label{eq_EHsoln}
\begin{align}
q^+ &= r\, \frac{f_2^+}{f_1^+} \Big(u^{\1+} - \frac{2\ri}{r^2} \,u_\2^- g^{++} \Big)~, \qquad
p^+ = r \frac{\tilde f_2^+}{\tilde f_1^+} \Big(u^{\2+} - \frac{2\ri}{r^2} \,u_\1^- g^{++} \Big)~, \\
f_a^\pm &:= f_a^i u_i^\pm~, \qquad \tilde f_a^\pm := \bar f_a^i u_i^{\pm}~, \quad a=1,2
\end{align}
\end{subequations}
where $f_a^+$ is constrained by
\begin{align}\label{eq_EHconstraint}
\xi^{++} = \ri\, f_1^+ \tilde f_1^+ + \ri \, f_2^+ \tilde f_2^+~,
\end{align}
and $g^{++}$ is a function of $\ell^{++}$, which is given by
\begin{align}
\ell^{++} = -\ri \, q^+ p^+ = -\ri\, f_2^+ \tilde f_2^+~.
\end{align}
As with the Taub-NUT solution, the product of $q^+$ and $p^+$ is simple and
$t$-independent,  $q^+ p^+ = f_2^+ \tilde f_2^+$.

The constrained complex coordinates $f_a^i$ correspond to those
originally discussed in \cite{CF, A-GF:EH}. These models were shown
to derive from a harmonic action involving two minimally
coupled hypermultipelts and an auxiliary vector multiplet in \cite{GIOT}
(based on the construction given in $\cN=1$ language \cite{RocekTownsend, LR:STDuality}),
which can be understood as the universal description of the Eguchi-Hanson
model in harmonic superspace. The version \eqref{eq_EHHarm} we have given here corresponds
to dualizing one of the hypermultiplets in \cite{GIOT} into an $(\omega, L^{++})$ system
and then applying the vector multiplet equation of motion, as discussed
in \cite{GIOS}.

The solution \eqref{eq_EHsoln} is valid only in a certain coordinate patch where
$|f_{1 \2}|^2 < |f_{1\1}|^2$. There is another solution found by
exchanging $f_1^+ \leftrightarrows f_2^+$, which is valid for 
$|f_{2 \2}|^2 < |f_{2\1}|^2$. These two patches cover the entire region
defined by \eqref{eq_EHconstraint}.
The reason two solutions are needed is apparent by noting that $q^+$
must be well-defined over the real $S^2$, but the first solution
has an apparent singularity when $\z$ is chosen so that $f_1^+$ vanishes,
that is, $\z = -f_{1 \1} / f_{1 \2}$. In the first coordinate patch,
this occurs for $|\z|>1$, and one can check that for this range, the
apparent singularity in $q^+$ is actually resolved: as
$\z \rightarrow -f_{1 \1} / f_{1 \2}$, one finds
$2\ri \,u_\2^- g^{++} \rightarrow u^{\1+}\, r^2$ and $q^+$ approaches
a finite value. Similarly, $p^+$ has an apparent singularity at the
antipodal point, which is resolved in the same fashion.

Now let us derive the Eguchi-Hanson model in projective superspace.
The on-shell projective Lagrangian is (after complexifying)
\begin{align*}
F^{++} &=
	\frac{\ri}{2} Q_0^+ P_0^+
	+ \frac{\ri}{2} Q_1^+ P_1^+
	+ \frac{\ri}{u^{\1+} u^{\2+}} \int_0^1 \rd t\,
	\Big(H^{+4} -\frac{1}{2} P^+ D^{++} Q^{+} + \frac{1}{2} Q^{+} D^{++} P^+ \Big) \eol
	&= -\ri f_2^+ \tilde f_2^+
	+ \frac{\ri}{u^{\1+} u^{\2+}} \int_0^1 \rd t\,
	\Big(H^{+4} -\frac{1}{2} P^+ \frac{\pa H^{+4}}{\pa P^+}
	- \frac{1}{2} Q^{+} \frac{\pa H^{+4}}{\pa Q^+} \Big) \eol
	&= -\ri f_2^+ \tilde f_2^+
	+ \frac{\ri}{u^{\1+} u^{\2+}} \int_0^1 \rd t\,
	\Big(\frac{2}{r^2} \frac{\xi^{++}}{\ri f_1^+ \tilde f_1^+} (g^{++})^2
	+ \frac{2}{r^2} D^{++} (\xi^{+-} g^{++}) \frac{f_2^+ \tilde f_2^+}{f_1^+ \tilde f_1^+}\Big)
\end{align*}
There are two $t$-integrals to do. The first is nearly identical
to the integral we performed which led to \eqref{eq_ImpTensDerived}. It yields
\begin{align}
\frac{\xi^{++}}{\ri f_1^+ \tilde f_1^+} \left(\ell^{++} - (\ell^{++} + \xi^{++})
	\ln (1+\ell^{++} / \xi^{++}) \right)
\end{align}
The second is a total derivative, which simplifies its evaluation:
\begin{align}
- \ell^{++} \frac{f_2^+ \tilde f_2^+}{f_1^+ \tilde f_1^+}~.
\end{align}
Combining all these terms and making use of the constraint \eqref{eq_EHconstraint},
we find
\begin{align}
F^{++} = -\xi^{++} \ln (1 + \ell^{++} / \xi^{++})~.
\end{align}

Now we must identify the arctic multiplet $\U^+$ and its conjugate $\breve\U^+$.
They are given by
\begin{align}\label{eq_EHprojSoln}
\U^+ = Q^+_0 = r\,u^{\1+}\frac{f_2^+}{f_1^+}~, \qquad
\breve \U^+ = P^+_1 = r\, u^{\2+} \frac{\tilde f_2^+}{\tilde f_1^+}~.
\end{align}
One can check that
\begin{align}
1 + \ell^{++} / \xi^{++} = \frac{1}{1+ \ri\, \U^+ \breve \U^+ / \xi^{++}}
\end{align}
which yields
\begin{align}
F^{++} = \xi^{++} \ln\left( 1 + \ri\, \U^+ \breve \U^+ / \xi^{++}\right)~.
\end{align}
This is indeed the projective action for Eguchi-Hanson \cite{GK1}.
Now let's check that it obeys the equations of motion. We observe that
\begin{align}
\Gamma^{+} := P_0^+ = r \frac{\tilde f_2^+}{\tilde f_1^+} \Big(u^{\2+} - \frac{\ri}{r^2 u^{\1+}} \ell^{++} \Big)
	= \frac{\breve \U^+}{1+ \ri\, \U^+ \breve\U^+ / \xi^{++}}
	= -\ri \frac{\pa F^{++}}{\pa \U^+}
\end{align}
is arctic. Similarly, $\breve\Gamma^+ := -Q_1^+$ is antarctic and
obeys \eqref{eq_ProjEOM}.

Finally, we note that the explicit solution for $\U^+$
\eqref{eq_EHprojSoln} can be shown to match that given originally in \cite{GK1}.
One introduces the variables $\Phi$ and $\Sigma$ defined as the first
two terms in the $\z$ expansion of $\U^+$:
\begin{align}
\U^{+} = u^{\1+} (\Phi + \z \Sigma + \cdots)~.
\end{align}
One finds $\Phi$ and $\Sigma$ are given by
\begin{align}
\Phi = r \frac{f_{2 \1}}{f_{1 \1}}~, \qquad
\Sigma = r \frac{f_{2 \2}}{f_{1\1}} (1 + \Phi \bar \Phi / r^2)~.
\end{align}
In terms of these variables, the solution for $\U^+$ becomes
\begin{align}
\U^+ = u^{\1+} \frac{\Phi + \z \Sigma / (1 + \Phi \bar \Phi / r^2)}{1 - \zeta \bar \Phi \Sigma / (r^2+\Phi\bar\Phi)}
\end{align}
in complete agreement with \cite{GK1}.

\section{Discussion}
For brevity's sake, we have given only the construction relating harmonic and
projective descriptions of nonlinear sigma models and two nontrivial examples here.
Many other nice examples, including several broad
classes, remain to be understood. One particular class of interest
is that of superconformal nonlinear sigma models: these are known in
both harmonic \cite{GIOS:Conformal} and in projective superspace \cite{K:5Dcomp, K:SPH}.
Another class is the set of nonlinear sigma models on tangent bundles
of Hermitian symmetric spaces, which have been fully analyzed from the
projective approach \cite{GK1, GK2, AN, AKL1, AKL2, KN}.

An important consequence of this construction is that it points toward a unifying
framework for the harmonic and projective descriptions of hyperk\"ahler
target spaces. It was shown in \cite{GIOS:HKgeo} that the harmonic
description of hyperk\"ahler geometry was universal: all hyperk\"ahler metrics
come from some harmonic nonlinear sigma model  (at least locally).
It was similarly argued in \cite{LR:Properties} that the same holds
for projective descriptions. It would be quite interesting to
see if one could ``complexify'' the geometric approach taken in \cite{GIOS:HKgeo}
to construct a projective analogue.

One interesting possibility that we have avoided addressing is whether
it might be possible to reconstruct harmonic actions and solutions
directly from projective solutions. On the one hand, this seems
impossible since information has been lost in moving from the harmonic
to the projective formulation. Indeed, this is the dominant point of
view \cite{K:DP, JS1}. However, the analogy to Hamiltonian mechanics
that we discussed in the introduction suggests an alternative.
If we view the symplectic action $F$ as a function of not just the endpoint
positions but also the endpoint \emph{times}, then the Hamiltonian can
be reconstructed by varying either of the endpoints,
\begin{align}
\frac{\pa F}{\pa t_1} = H(q_1, p_1, t_1)~, \qquad
\frac{\pa F}{\pa t_0} = - H(q_0, p_0, t_0)~.
\end{align}
Certainly we cannot vary the ``times'' in the complexified $S^2$ since
the boundary values of $t$ correspond to something meaningful -- the two poles.
However, it might be possible to vary the location of the poles themselves.
Presumably, a single harmonic action might then correspond to a family of
projective actions. This conjecture probably exceeds
the elasticity of the analogy, but it may be interesting to investigate further.

\vspace{0.5cm}

\noindent
{\bf Acknowledgements:}\\
The author would like to thank Sergei Kuzenko for invaluable discussions, for reading
the manuscript, and for suggesting numerous improvements.
This work is supported by the Australian Research Council (grant No. DP1096372).

\appendix

\section{The complexified $S^2$}\label{app_cS2}
The two-sphere $S^2$ is defined as the set of points $X^I \in \mathbb R^3$ obeying
$\sum_I (X^I)^2 = 1$. It can be related to the space $\mathbb CP^1$, with homogeneous
coordinate $v^i$, via the identification $X^I = \bar v_j (\sigma^I)^j{}_k v^k / (v, \bar v)$
where $\sigma^I$ are the usual Pauli matrices.
This defines $v^i$ up to the equivalence relation $v^i \sim c v^i$ for complex $c \neq 0$.

This space is straightforwardly complexified\footnote{The author would like to thank
Sergei Kuzenko for educating him about the complexified $S^2$.} to the standard affine quadric $Q^2$, defined
as (see \textit{e.g.} \cite{Feix})
\begin{align}
Q^2 = \Big\{Z^I \in \mathbb C^3 : \sum_I (Z^I)^2 = 1\Big\}~.
\end{align}
One can check that $Q^2 \subset \mathbb CP^1 \times \mathbb CP^1$ by identifying
\begin{align}
Z^I = \frac{1}{(v, \bar w)} \, \bar w_j (\sigma^I)^j{}_k v^k~, \qquad (v, \bar w) := v^k \bar w_k~.
\end{align}
This defines $v^i$ and $\bar w_j$ up to the identifications
\begin{align}
v^i \sim c v^i~, \qquad \bar w_j \sim \bar d \bar w_j~, \qquad c, \bar d \in \mathbb C - \{0\}~.
\end{align}
$Q^2$ can be identified with $\mathbb CP^1 \times \mathbb CP^1$ with the region $(v,\bar w) = 0$ removed.

This complexified two-sphere obviously has a real $S^2$ as a compact subspace with $Z^I = X^I$,
or $v^i \propto w^i$:
\begin{align}\label{eq_compS^2}
S^2 = \Big\{(v^i, \bar w_i) \in \mathbb CP^1 \times \mathbb CP^1 : v^i \propto w^i \Big\}~.
\end{align}
The standard integral on the $S^2$ is \eqref{eq_intS2}, which we can interpret as
\begin{align}
\int_{S^2} \rd u = \frac{1}{2\pi\ri} \int_{S^2} \frac{v^i \rd v_i \wedge \bar w_j \rd \bar w^j}{(v,\bar w)^2}~,
\end{align}
for $S^2 \subset Q^2$ defined by \eqref{eq_compS^2}.

Now suppose we have a scalar function $F$ defined on $S^2$. It possesses a convergent
harmonic expansion \eqref{eq_Fweight0}. $F$ may be analytically continued to a bi-holomorphic function on $Q^2$
\begin{align}\label{eq_Fexpansion}
F = f + f^{(ij)} \frac{v_i \bar w_j}{(v, \bar w)} + \cdots
\end{align}
which in general only converges in the vicinity of $v\propto w$.
Along $S^2$, we know that
\begin{align}\label{eq_intF}
\int_{S^2} \rd u \,F = f~.
\end{align}
We would like to extend this result to a different surface on $Q^2$.
Let us introduce the complex coordinates $\z$ and $t$ given by 
\begin{align}
v^i = v^\1 (1, \z)~, \qquad \bar w_i = \frac{\bar w_\1}{1-t} (1-t, t / \z)~.
\end{align}
In terms of these coordinates,
\begin{align}
\int_{S^2} \rd u \,F = \int_{S^2} \frac{\rd \zeta \wedge \rd t}{2\pi \ri \zeta} \,F
\end{align}
where $S^2$ can now be identified as the region
\begin{align}
S^2 = \Big\{(\z, t) : 0 \leq t \leq 1~, |\z|^2 = \frac{t}{1-t} \Big\}~.
\end{align}
Now we keep the domain of $t$ fixed but deform the $\z$ contour
to $|\z|=1$. We must assume that the expansion \eqref{eq_Fexpansion}
remains valid after the deformation. Let $\cP^2$ be the corresponding
region with $|\z|=1$. To prove that \eqref{eq_intF} holds, we must prove that the harmonic result
\begin{align}
\int_{S^2} \rd u \,u^{i_1+} \cdots u^{i_n+} u_{j_1}^- \cdots u_{j_n}^-=
	\frac{n!}{(n+1)!} \delta^{(i_1}_{(j_1} \cdots \delta^{i_n)}_{j_n)}
\end{align}
holds when we replace $S^2$ with $\cP^2$, for which
\begin{align}
u^{i+} = u^{\1+} (1, \z)~, \qquad u_{j}^- = \frac{1}{u^{\1+}} (1-t, t/\z)~.
\end{align}
This is quite straightforward. We first observe that the $\z$ contour
integral is non-vanishing only for $\z$-independent harmonic products. This implies
that we must have the same number of $u^{\2+}$ and $u_\2^-$ factors,
which implies the Kronecker delta structure.
So we are reduced to checking for the non-vanishing case of $p$ $u^{\1+}$'s and
$n-p$ $u^{\2+}$'s:
\begin{align}
\int_{\cP^2} \rd u \,(u^{\1+} u_{\1}^-)^p (u^{\2+} u_{\2}^-)^{n-p}  = \frac{p! \,(n-p)!}{(n+1)!}
\end{align}
Performing the $\z$ integral, we find
\begin{align}
\int_0^1 \rd t\, (1-t)^p t^{n-p} = \frac{p! \,(n-p)!}{(n+1)!}
\end{align}
which does indeed hold.
So we are justified in performing the continuation of $S^2$ to $\cP^2$
(dubbed a ``Wick rotation'' in \cite{JS1}) provided that the integrand still
possesses a convergent expansion.

Finally, we should comment briefly about what the surface $\cP^2$ looks like
as a subspace of $Q^2 \subset \mathbb CP^1 \times \mathbb CP^1 = S^2 \times S^2$.
It is not hard to see that
\begin{align}
\cP^2 = \Big\{(\theta_1, \phi_1, \theta_2, \phi_2) \in S^2 \times S^2 : \phi_1=\phi_2~, \theta_1 = \pi/2\Big\}~.
\end{align}
In other words, $\cP^2$ arises when we identify the azimuthal angles on the two $S^2$'s
and restrict to the equator of the first $S^2$. In particular,
$\cP^2$ is a compact surface.


\begin{thebibliography}{66}

\bibitem{Zumino:1979et} 
  B.~Zumino,
  ``Supersymmetry and K\"ahler manifolds,''
  Phys.\ Lett.\ B {\bf 87}, 203 (1979).

\bibitem{A-GF}
L.~Alvarez-Gaum\'e and D.~Z.~Freedman,
``Geometrical structure and ultraviolet finiteness in 
the supersymmetric sigma model,''
Commun.\ Math.\ Phys.\  {\bf 80}, 443 (1981);
  ``Potentials for the supersymmetric nonlinear sigma model,''
  Commun.\ Math.\ Phys.\  {\bf 91}, 87 (1983).
  
\bibitem{CF}
  T.~L.~Curtright and D.~Z.~Freedman,
  ``Nonlinear sigma models with extended supersymmetry in four-dimensions,''
  Phys.\ Lett.\  B {\bf 90}, 71 (1980)
  [Erratum-ibid.\  B {\bf 91}, 487 (1980)].


\bibitem{GIKOS}
A.~Galperin, E.~Ivanov, S.~Kalitsyn, V.~Ogievetsky and E.~Sokatchev,
``Unconstrained N = 2 matter, Yang-Mills and supergravity theories in harmonic
superspace,''
Class.\ Quant.\ Grav.\  {\bf 1}, 469 (1984).

\bibitem{GIOS} 
  A.~S.~Galperin, E.~A.~Ivanov, V.~I.~Ogievetsky and E.~S.~Sokatchev,
  ``Harmonic superspace,''
  Cambridge, UK: Univ. Pr. (2001) 306 p

\bibitem{KLR}
A. Karlhede, U. Lindstr\"om and M. Ro\v cek,
``Self-interacting tensor multiplets in N = 2 superspace,''
Phys.\ Lett.\ B {\bf 147}, 297 (1984). 


\bibitem{LR:Projective}
U.~Lindstr\"om and M.~Ro\v{c}ek,
``New hyperk\"ahler  metrics  and new supermultiplets,''
 Commun.\ Math.\ Phys.\  {\bf 115}, 21 (1988);

\bibitem{LR:SYM}
U.~Lindstr\"om and M.~Ro\v{c}ek,
   ``N = 2 super Yang-Mills theory in projective superspace,''
Commun.\ Math.\ Phys.\  {\bf 128}, 191 (1990).


\bibitem{K:DP} 
  S.~M.~Kuzenko,
  ``Projective superspace as a double punctured harmonic superspace,''
  Int.\ J.\ Mod.\ Phys.\ A {\bf 14}, 1737 (1999)
  [hep-th/9806147].

\bibitem{JS1}
  D.~Jain and W.~Siegel,
  ``Deriving projective hyperspace from harmonic,''
  Phys.\ Rev.\ D {\bf 80}, 045024 (2009)
  [arXiv:0903.3588 [hep-th]].

\bibitem{JS2} 
  D.~Jain and W.~Siegel,
  ``On projective hoops: loops in hyperspace,''
  Phys.\ Rev.\ D {\bf 83}, 105024 (2011)
  [arXiv:1012.3758 [hep-th]].

\bibitem{JS3} 
  D.~Jain and W.~Siegel,
  ``N=2 super-Yang-Mills theory from a Chern-Simons action,''
  arXiv:1203.2929 [hep-th].

\bibitem{GIOS:HMHS} 
  A.~Galperin, E.~Ivanov, V.~Ogievetsky and E.~Sokatchev,
  ``Hyperkahler metrics and harmonic superspace,''
  Commun.\ Math.\ Phys.\  {\bf 103}, 515 (1986).

\bibitem{GIOS:HKgeo}
  A.~S.~Galperin, E.~A.~Ivanov, V.~I.~Ogievetsky and E.~Sokatchev,
  ``Gauge field geometry from complex and harmonic analyticities II. Hyper-kahler case,''
  Annals Phys.\  {\bf 185}, 22 (1988).

\bibitem{GIO:Duality} 
  A.~S.~Galperin, E.~A.~Ivanov and V.~I.~Ogievetsky,
  ``Duality transformations and most general matter self-couplings in N=2 supersymmetry,''
  Nucl.\ Phys.\ B {\bf 282}, 74 (1987).

\bibitem{GO:Hamiltonian} 
  A.~Galperin and V.~Ogievetsky,
  ``N=2 D = 4 supersymmetric sigma models and Hamiltonian mechanics,''
  Class.\ Quant.\ Grav.\  {\bf 8}, 1757 (1991).



\bibitem{RoslySchwarz}
A.~A.~Rosly and A.~S.~Schwarz,
  ``Supersymmetry in a space with auxiliary dimensions,''
 Commun.\ Math.\ Phys.\  {\bf 105}, 645 (1986).

\bibitem{Rosly}
 A.~A.~Rosly,
``Super Yang-Mills  constraints 
as integrability conditions,'' in {\it Proceedings of the International 
Seminar on Group Theoretical 
Methods in Physics},'' (Zvenigorod, USSR, 1982),
M. A. Markov  (Ed.), 
Nauka, Moscow, 1983, Vol. 1, p. 263 (in Russian)~.

\bibitem{K:Lectures} 
  S.~M.~Kuzenko,
  ``Lectures on nonlinear sigma-models in projective superspace,''
  J.\ Phys.\ A A {\bf 43}, 443001 (2010)
  [arXiv:1004.0880 [hep-th]].

\bibitem{GonzalezRey:1} 
  F.~Gonzalez-Rey, M.~Rocek, S.~Wiles, U.~Lindstrom and R.~von Unge,
  ``Feynman rules in N=2 projective superspace: 1. Massless hypermultiplets,''
  Nucl.\ Phys.\ B {\bf 516}, 426 (1998)
  [hep-th/9710250].


\bibitem{GK1}
S.~J.~Gates Jr. and S.~M.~Kuzenko,
``The CNM-hypermultiplet nexus,''
Nucl.\ Phys.\ B {\bf 543}, 122 (1999) [arXiv:hep-th/9810137].

\bibitem{GK2}
S.~J.~Gates Jr. and S.~M.~Kuzenko,
``4D N = 2 supersymmetric off-shell sigma models on the cotangent  
bundles of  K\"ahler manifolds,'' Fortsch.\ Phys.\  {\bf 48}, 115 (2000)
[arXiv:hep-th/9903013].

\bibitem{LR:Properties} 
  U.~Lindstr\"om and M.~Ro\v cek,
  ``Properties of hyperkahler manifolds and their twistor spaces,''
  Commun.\ Math.\ Phys.\  {\bf 293}, 257 (2010)
  [arXiv:0807.1366 [hep-th]].


\bibitem{K:5Dcomp}
  S.~M.~Kuzenko,
  ``On compactified harmonic/projective superspace, 5-D superconformal theories, and all that,''
  Nucl.\ Phys.\ B {\bf 745}, 176 (2006)
  [hep-th/0601177].

\bibitem{K:SPH} 
  S.~M.~Kuzenko,
  ``On superconformal projective hypermultiplets,''
  JHEP {\bf 0712}, 010 (2007)
  [arXiv:0710.1479 [hep-th]].


\bibitem{AN}
M.~Arai and M.~Nitta,
``Hyper-K\"ahler sigma models on (co)tangent bundles with SO(n) isometry,''
Nucl.\ Phys.\ B {\bf 745}, 208 (2006)  [arXiv:hep-th/0602277].


\bibitem{AKL1}
  M.~Arai, S.~M.~Kuzenko and U.~Lindstr\"om,
  ``Hyperk\"ahler sigma models on cotangent bundles of Hermitian symmetric
  spaces using projective superspace,''
  JHEP {\bf 0702}, 100 (2007)  [arXiv:hep-th/0612174].
  
\bibitem{AKL2}
  M.~Arai, S.~M.~Kuzenko and U.~Lindstr\"om,
  ``Polar supermultiplets, Hermitian symmetric spaces and hyperk\"ahler
  metrics,''  JHEP {\bf 0712}, 008 (2007)
 [arXiv:0709.2633 [hep-th]].


\bibitem{KN} 
  S.~M.~Kuzenko and J.~Novak,
  ``Chiral formulation for hyperkahler sigma-models on cotangent bundles of symmetric spaces,''
  JHEP {\bf 0812}, 072 (2008)
  [arXiv:0811.0218 [hep-th]].



\bibitem{Kuzenko:SigmaDuality} 
  S.~M.~Kuzenko,
  ``N = 2 supersymmetric sigma-models and duality,''
  JHEP {\bf 1001}, 115 (2010)
  [arXiv:0910.5771 [hep-th]].

\bibitem{K:CommentsProj} 
  S.~M.~Kuzenko,
  ``Comments on N = 2 supersymmetric sigma models in projective superspace,''
  J.\ Phys.\ A A {\bf 45}, 095401 (2012)
  [arXiv:1110.4298 [hep-th]].

\bibitem{HiKLR} 
  N.~J.~Hitchin, A.~Karlhede, U.~Lindstrom and M.~Rocek,
  ``Hyperkahler metrics and supersymmetry,''
  Commun.\ Math.\ Phys.\  {\bf 108}, 535 (1987).


\bibitem{GIOS:Conformal} 
  A.~Galperin, E.~Ivanov, V.~Ogievetsky and E.~Sokatchev,
  ``Conformal invariance in harmonic superspace,''
  in *Batalin, I.A. (ed.) et al.: Quantum Field Theory and Quantum Statistics, Vol. 2*, 233-248
  and Dubna JINR - 85-363 (85,Rec.Jul.) 18p

\bibitem{GIO:Quat} 
  A.~Galperin, E.~Ivanov and O.~Ogievetsky,
  ``Harmonic space and quaternionic manifolds,''
  Annals Phys.\  {\bf 230}, 201 (1994)
  [hep-th/9212155].

\bibitem{GIO:Tensor} 
  A.~Galperin, E.~Ivanov and V.~Ogievetsky,
  ``Superspace actions and duality transformations for N=2 tensor multiplets,''
  Sov.\ J.\ Nucl.\ Phys.\  {\bf 45}, 157 (1987)
  [Yad.\ Fiz.\  {\bf 45}, 245 (1987)]
  [Phys.\ Scripta T {\bf 15}, 176 (1987)].

\bibitem{Gibbons:1987pk} 
  G.~W.~Gibbons, D.~Olivier, P.~J.~Ruback and G.~Valent,
  ``Multicenter metrics and harmonic superspace,''
  Nucl.\ Phys.\ B {\bf 296}, 679 (1988).

\bibitem{GIOT} 
  A.~Galperin, E.~Ivanov, V.~Ogievetsky and P.~K.~Townsend,
  ``Eguchi-Hanson type metrics from harmonic superspace,''
  Class.\ Quant.\ Grav.\  {\bf 3}, 625 (1986).


\bibitem{A-GF:EH} 
  L.~Alvarez-Gaume and D.~Z.~Freedman,
  ``Ricci Flat Kahler manifolds and supersymmetry,''
  Phys.\ Lett.\ B {\bf 94}, 171 (1980).


\bibitem{RocekTownsend} 
  M.~Rocek and P.~K.~Townsend,
  ``Three foop finiteness of the N=4 supersymmetric nonlinear sigma model,''
  Phys.\ Lett.\ B {\bf 96}, 72 (1980).

\bibitem{LR:STDuality} 
  U.~Lindstrom and M.~Rocek,
  ``Scalar tensor duality and N=1, N=2 nonlinear sigma models,''
  Nucl.\ Phys.\ B {\bf 222}, 285 (1983).


\bibitem{Feix}
B. Feix, ``Hyperk\"ahler metrics on  cotangent bundles,''
Cambridge PhD thesis, 1999; ``Hyperk\"ahler metrics on  cotangent bundles,''
J. Reine Angew. Math. {\bf 532}, 33 (2001).



\end{thebibliography}
\end{document}